\documentclass[pre,showpacs,twocolumn,floatfix,bibnotes]{revtex4}
\usepackage{here}
\usepackage[dvips]{graphicx}

\newcounter{Fig}

\begin{document}

\begin{sloppy}

\title{Chaos suppression in the parametrically driven Lorenz system}

\author{Chol-Ung Choe$^{1,2}$, Hartmut Benner$^1$, and Yuri S.
Kivshar$^{3}$}

\affiliation{$^1$Institut f\"ur Festk\"orperphysik, Technische
Universit\"at Darmstadt, D-64289 Darmstadt, Germany\\
$^2$Department of Physics, University of Science, Pyongyang, DPR
Korea
\\  $^3$ Nonlinear Physics Center, Research School of Physical
Sciences and Engineering, Australian National University, Canberra
ACT 0200, Australia}

\begin{abstract}
We predict theoretically and verify experimentally the suppression
of chaos in the Lorenz system driven by a high-frequency periodic
or stochastic parametric force. We derive the theoretical criteria
for chaos suppression and verify that they are in a good agreement
with the results of numerical simulations and the experimental
data obtained for an analog electronic circuit.
\end{abstract}


\maketitle

\section{Introduction}

Control of the chaotic dynamics of complex nonlinear systems is
one of the most important and rapidly developing topics in applied
nonlinear science. In particular, the concept of chaos control,
first introduced in Ref.~\cite{ref_5}, has attracted a great deal
of attention over the last decade. Among different methods for
controlling chaos that have been suggested by now, the so-called
non-feedback control is attractive because of its simplicity: no
measurements or extra sensors are required. The idea of this
method is to change the complex behavior of a nonlinear stochastic
system by applying a properly chosen external force. It is
especially advantageous for ultra-fast processes, e.g., at the
molecular or atomic level, where no possibility to measure the
system variables exists.

Many of the suggested non-feedback methods employ the external
forces acting at the system frequencies~\cite{ref_7}, including
parametric perturbations with the frequency that is in resonance
with the main driving force. In particular, changing the phase and
frequency of a parameter perturbation in a bistable mechanical
device was shown to either decrease or increase the threshold of
chaos~\cite{ref_8}. Similarly, unstable periodic orbits are known
to be stabilized by a low-frequency modulation of a system control
parameter~\cite{ref_12}, when the control frequency is much lower
than the system frequency.

A possibility to change significantly the system dynamics by
applying a high- (rather than low- ) frequency force is known for
almost a century. As a textbook example, we mention the familiar
stabilization of a reverse pendulum (known as the Kapitza
pendulum) by rapid vertical oscillations of its
pivot~\cite{ref_13}. This discovery triggered the development of
vibrational mechanics~\cite{ref_14} where the general analysis of
nonlinear dynamics in the presence of rapidly varying forces is
based on the Krylov-Bogoljubov averaging method~\cite{ref_15}. In
the control theory, the high-frequency forces and parameter
modulations are usually used for the vibrational control of
non-chaotic nonlinear systems~\cite{ref_16}. However, as was shown
for the example of the Duffing oscillator~\cite{ref_17,ref_exp},
chaos suppression can also be achieved by applying a
high-frequency parametric force. Later, chaos suppression in the
Belousov-Zhabotinsky reaction was demonstrated numerically by
adding white noise~\cite{ref_18}, and the effect of random
parametric noise on a periodically driven damped nonlinear
oscillator was studied by the Melnikov analysis~\cite{ref_19}.

In this paper we apply the concepts of the nonresonant
non-feedback control to the Lorenz system and demonstrate
analytically, numerically, and also experimentally that the
suppression of the chaotic dynamics can be achieved by applying a
high-frequency parametric or random parametric force. The Lorenz
system, found in 1963, is known to produce the canonical chaotic
attractor in a simple three-dimensional autonomous
system~\cite{ref_1,ref_2}, and it can be applied to describe many
interesting nonlinear systems, ranging from thermal
convection~\cite{ref_3} to lasers dynamics~\cite{ref_4}.

The paper is organized as follows. In Sec. II we present our model
and outline the theoretical method and results. By applying the
averaging method, we derive the effective Lorenz equations with
the renormalized control parameter and obtain the conditions for
chaos suppression. In Sec. III, we demonstrate the suppression of
chaos by means of direct numerical simulations and also present
the experimental data obtained for an analog electronic circuit.
Finally, Sec. IV concludes the paper.

\section{Theoretical approach}

\subsection{Model}

We consider the familiar Lorenz system~\cite{ref_1} driven by a
parametric force. In the dimensionless variables, the nonlinear
dynamics is governed by the equations
\begin{equation}
   \begin{array}{l} {\displaystyle
       \dot{x} = \sigma (y-x),
   } \\*[9pt] {\displaystyle
      \dot{y} = r [1 +f(t)]x - y -xz,
       } \\*[9pt] {\displaystyle
      \dot{z} = xy-bz,
   }\end{array}
   \label{eq_1}
\end{equation}
where the dots stand for the derivatives in time, $\sigma$, $r$,
and $b$ are the parameters of the Lorenz model, and the function
$f(t)$ describes a parametric force. For definiteness but without
restrictions of generality, we select the standard set of the
parameter values, $\sigma =10$ and $b=8/3$, whereas the parameter
$r$ is assumed to vary. It is well known~\cite{ref_1} that, in the
absence of the parametric driving force $f(t)$, the original
Lorenz equations demonstrate different dynamical regimes on
variation of the control parameter $r$, which are associated with
the existence and stability of several equilibrium states. In
brief, the system dynamics can be characterized by three regimes:
\begin{itemize}

\item the $G_1$ regime, for $r <1$: there exists the only stable
fixed point at the origin, $(x, y, z)=(0,0,0)$;

\item
the $G_2$ regime, for $1< r <24.74$: the origin becomes unstable,
two new fixed points appear, $(x, y, z)= (\pm\sqrt{b(r-1)},
\pm\sqrt{b(r-1)}/r, r-1)$;

\item
the $G_3$ regime, for $r>24.74$: no stable fixed points exist,
chaotic dynamics occur with a strange attractor.

\end{itemize}

We consider the special case of the general model~(\ref{eq_1}),
assuming that the characteristic frequency of the parametric force
$f(t)$ is much larger than the characteristic frequency of the
unforced Lorenz system, where the frequency (which is the
mean-time derivative of the phase) of the Lorenz system can be
defined as~\cite{ref_10}
\begin{equation}
\label{eq_2}
 \omega_0 = \lim_{T \rightarrow \infty} \frac{2\pi
N(T)}{T},
\end{equation}
%                                                                        ,
where $N(T)$ is the number of turns performed in $T$.

\subsection{Periodic driving force}

First, we consider the case of {\em a periodic driving force},
\begin{equation}
\label{eq_3}
 f(t) = k \cos (\omega t).
\end{equation}
Assuming that the frequency of parametric modulations is large
($\omega \gg \omega_0$), we apply an asymptotic analytical
method~\cite{ref_13,ref_14} based on a separation of different
time scales, and derive the effective equations that describe the
slowly varying dynamics. To do this, we decompose every variable
into a sum of slowly and rapidly varying parts, i.e.,
\begin{equation}
\label{eq_4}
 x = X + \xi, \; y = Y + \eta, \; z = Z +\zeta,
\end{equation}
where the functions $\xi(t)$, $\eta(t)$, and $\zeta(t)$ describe
fast oscillations around the slowly varying envelope functions
$X(t)$, $Y(t)$, and $Z(t)$, respectively. The rapidly oscillating
corrections are assumed to be small in comparison with the slowly
varying parts, and their mean values during an oscillation period
vanish, i.e. $<x> = X$, $<y> =Y$, and $<z> =Z$. Substituting the
Eq.~(\ref{eq_4}) into Eq.~(\ref{eq_1}) with the force
(\ref{eq_3}), and averaging over the oscillation period, we obtain
the following equation,
\begin{equation}
   \begin{array}{l} {\displaystyle
       \dot{X} = \sigma (Y-X),
   } \\*[9pt] {\displaystyle
      \dot{Y} = rX -Y -XZ + rk < \xi \cos (\omega t)>,
       } \\*[9pt] {\displaystyle
      \dot{Z} = XY - bZ,
   }\end{array}
   \label{eq_5}
\end{equation}
where the terms $<\xi \zeta>$ and $<\xi \eta>$ are neglected
because, for large $\omega$, they are all of the higher orders in
$k\omega^{-1}$. Using Eq.~(\ref{eq_5}) and keeping only the terms
not smaller than those of the order of $k\omega^{-1}$ and $k$,
respectively, we find the equations,\ $\dot{\xi} = \sigma \eta$
and $\dot{\eta} = rk X \cos (\omega t)$. Regarding the function
$X$ as constant during the period of the function $f(t)$, we
obtain the solution $\xi = -\sigma r (k/\omega^2) X \cos (\omega
t)$, and thus $< \xi \cos(\omega t)> = -\sigma r (k/2\omega^2) X$.
Therefore, the averaged equations (\ref{eq_5}) are
\begin{equation}
   \begin{array}{l} {\displaystyle
       \dot{X} = \sigma (Y-X),
   } \\*[9pt] {\displaystyle
      \dot{Y} = r_{\rm eff}X -Y -XZ,
       } \\*[9pt] {\displaystyle
      \dot{Z} = XY - bZ,
   }\end{array}
   \label{eq_8}
\end{equation}
where
\begin{equation}
\label{eq_r1} r_{\rm eff} = r (1 -rK_{\omega}), \;\;\;\;
K_{\omega} = \frac{\sigma k^2}{2 \omega^2}>0.
\end{equation}
As a result, the averaged dynamics of the Lorenz system in the
presence of a rapidly varying parametric force is described by the
effective renormalized Lorenz equations (\ref{eq_8}) with the
effective control parameter $r_{\rm eff}$, so that all dynamical
regimes and the route to chaos discussed earlier can be applicable
directly to Eq. (\ref{eq_8}), assuming the effect of the
renormalization.

Therefore, in terms of the averaged system, for $r_{\rm eff} \leq
1$, i.e. under the condition
\begin{equation}
\label{eq_9}
 K_{\omega} \geq (r-1)/r^2,
\end{equation}
the fixed point at the origin remains stable and, therefore, in
terms of the original Lorenz system, the chaotic dynamics should
be suppressed. Next,  for $1 < r_{\rm eff} \leq 24.74$, i.e. under
the condition
\begin{equation}
\label{eq_10}
 (r - 24.74)/r^2 \leq K_{\omega} < (r-1)/r^2,
\end{equation}
the Lorenz chaos is also suppressed and the stable saddle-focus
points appears: $X_0 = Y_0 = \pm \sqrt{b(r_{\rm eff} -1)}$, $Z_0
=r_{\rm eff} -1$.

Relations (\ref{eq_9}) and (\ref{eq_10}) that follow from the
averaged equations, define the conditions for the chaos
suppression; they can be expressed as relations between the
amplitude and frequency of the rapidly varying oscillations,
\begin{equation}
\label{eq_11} k \geq \omega \sqrt{2(r-1)/\sigma r^2},
\end{equation}
\begin{equation}
\label{eq_12}
 \omega \sqrt{2(r-24.7)/\sigma r^2} \leq k < \omega
\sqrt{2(r-1)\sigma r^2}.
\end{equation}
The dependencies $K_{\omega} \sim r$ (see Eqs.~(\ref{eq_9}) and
(\ref{eq_10})) and $k \sim \omega$ (see Eqs.~(\ref{eq_11}) and
(\ref{eq_12})) are the key characteristics of the chaos
suppression effect that we verify numerically and compare with the
experimental data obtained with an analog electronic circuit (see
Figs.~4 and 5 below).

\subsection{Random driving force}

Now we turn to the case of a {\em random force}, and treat the
function $f(t)$ in Eq.~(\ref{eq_1}) as random by formally writing
$f(t) = \epsilon(t)$, where $\epsilon(t)$ describes a
bandwidth-limited noise with a power spectral density
\begin{equation}  \label{eq:1}
P_{\epsilon}(\omega) = \Biggr\{
   \begin{array}{l} {\displaystyle
      p(\omega),  \,\;\; {\rm when} \; \omega_1 < |\omega| < \omega_2,
   } \\*[9pt] {\displaystyle
    0, \,\;\; {\rm when} \; |\omega| <\omega_1, |\omega| > \omega_2,
   } \end{array}
   \label{eq_00}
\end{equation}
and the zero mean value, $<\epsilon(t)> =0$. In order to apply the
analytical method discussed above, the noise $\epsilon(t)$ is
assumed to be composed of high-frequency components only, i.e.
$\tilde{\omega} \gg \omega_0$, where
\[ \tilde{\omega} =\int^{\infty}_{-\infty} \omega P_{\epsilon} (\omega) d\omega
\left[ \int^{\infty}_{-\infty} P_{\epsilon}(\omega) d\omega
\right]^{-1}
\]
is the characteristic frequency of the bandwidth-limited
parametric fluctuations. Again, decomposing the system variables
into sums of slow and rapidly varying functions and averaging over
the period $T_0$, we obtain
\begin{equation}
   \begin{array}{l} {\displaystyle
       \dot{X} = \sigma (Y-X),
   } \\*[9pt] {\displaystyle
      \dot{Y} = rX -Y -XZ + rk < \xi(t) \epsilon(t)>,
       } \\*[9pt] {\displaystyle
      \dot{Z} = XY - bZ,
   }\end{array}
   \label{eq_14}
\end{equation}
where the value of $T_0$ is much smaller than the characteristic
time scale of the Lorenz system oscillations but sufficiently
larger than the time scale of fluctuations, i.e.
$2\pi/\tilde{\omega} < T_0 \ll 2\pi/\omega_0$. Then, the equations
for the rapid parts can be reduced to the equations, $\dot{\xi} =
\sigma \eta(t)$ and  $\dot{\eta} = r \epsilon(t) X$. Thus, the key
averaged quantity  becomes $<\xi(t) \epsilon(t)> = <\xi (t)
\ddot{\xi}(t) > /(\sigma r X)$. On the other hand,  for the
stationary random process $\xi(t)$, the following expressions are
valid,
\[
<\xi(t) \ddot{\xi(t)}> = - <\dot{\xi}^2(t)> = -
\int^{\infty}_{-\infty} \omega^2 P_{\xi}(\omega) d\omega,
\]
where $P_{\xi}(\omega) = |H(\omega)|^2 P_{\epsilon}(\omega)$ is
the power spectral density of $\xi(t)$ and $H(\omega) =-\sigma r
X/\omega^2$ is the frequency response function for the equation
$\ddot{\xi}= \sigma r X \epsilon(t)$. As a result, we obtain the
averaged equations in the form (\ref{eq_8}) where this time [cf.
Eq.~(\ref{eq_r1})]
\begin{equation}
\label{eq_r2}
 r_{\rm eff} = r(1-rK_{\epsilon}), \;\;\;
K_{\epsilon} = 2\sigma \int^{\omega_2}_{\omega_1}
\frac{p(\omega)}{\omega^2} d\omega.
\end{equation}
The quantity $K_{\epsilon}$ is related to the intensity of the
parametric fluctuations $\epsilon(t)$ and it is always positive.
Since the averaged equations have the same form as the original
unforced Lorenz equations but with the effective control parameter
$r_{\rm eff}$ instead of $r$, there appears the parameter region
where the chaotic dynamics is suppressed. In particular, for
$r_{\rm eff} \leq 1$, i.e., under the condition
\begin{equation}
\label{eq_18}
 K_{\epsilon} \geq (r-1)/r^2,
\end{equation}
the fixed point at the origin remains stable in the presence of
rapidly varying fluctuations. When $1 < r_{\rm eff} \leq 24.74$,
i.e. for
\begin{equation}
\label{eq_19} (r-24.74)/r^2 \leq K_{\epsilon} < (r-1)/r^2,
\end{equation}
two new saddle-focus fixed points appear, with the coordinates
$X_0 = Y_0 = \pm \sqrt{b(r_{\rm eff}-1)}$, $Z_0= r_{\rm eff} -1$.

%%%%%%%%%%%%%%%%%%%%%fig1
\begin{figure}
\centerline{\includegraphics[width=2.7in]{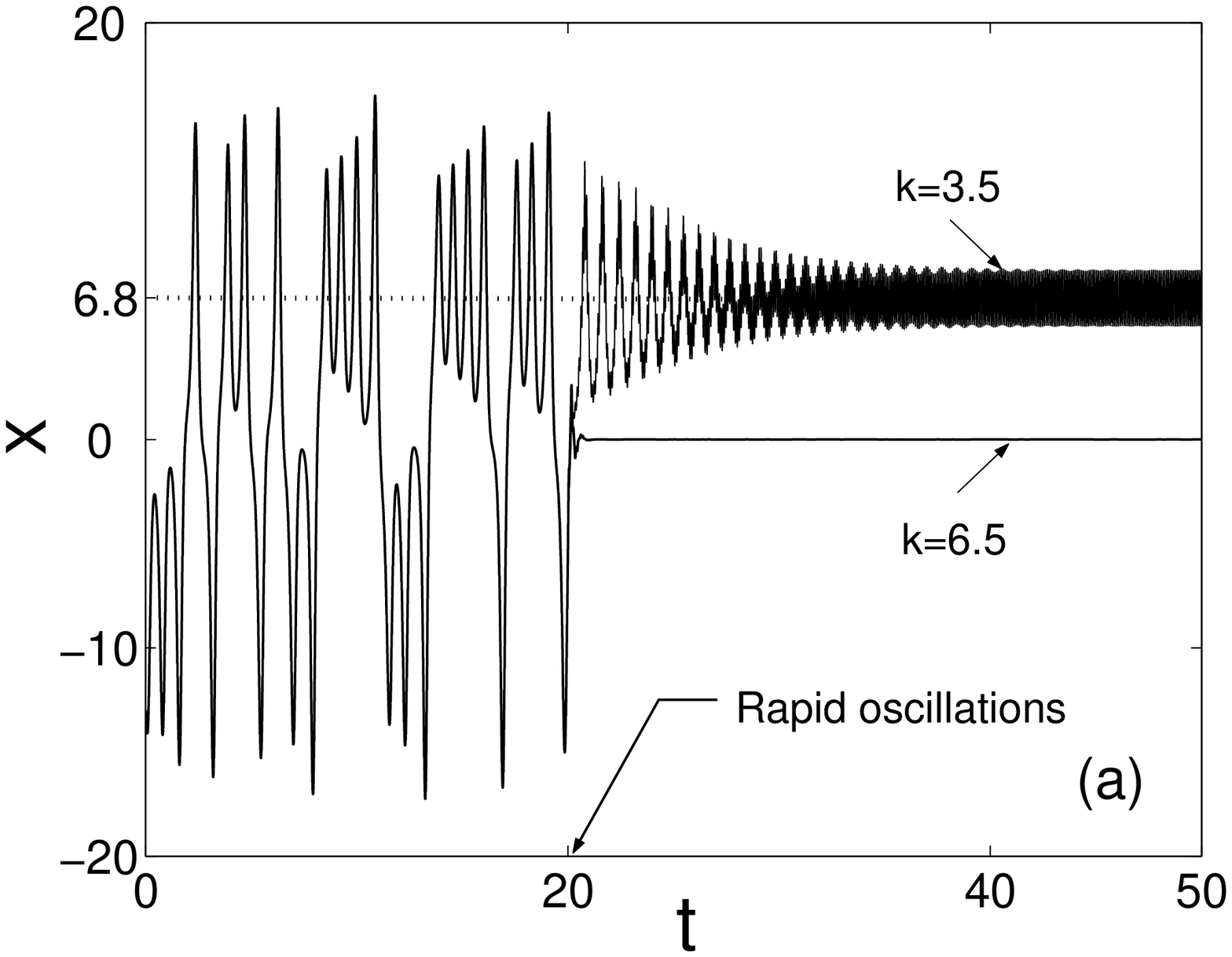}}
\centerline{\includegraphics[width=1.6in]{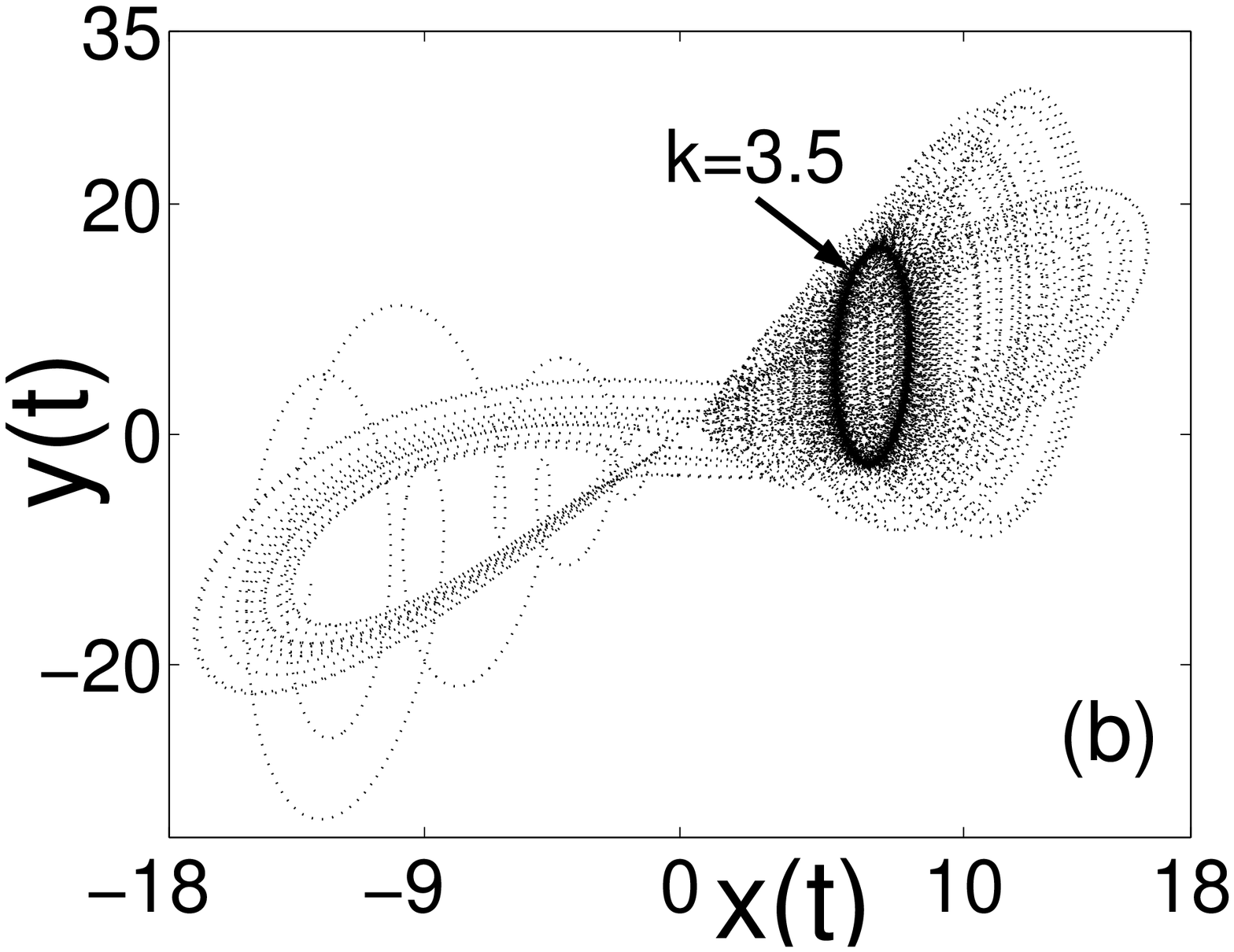}
\includegraphics[width=1.6in]{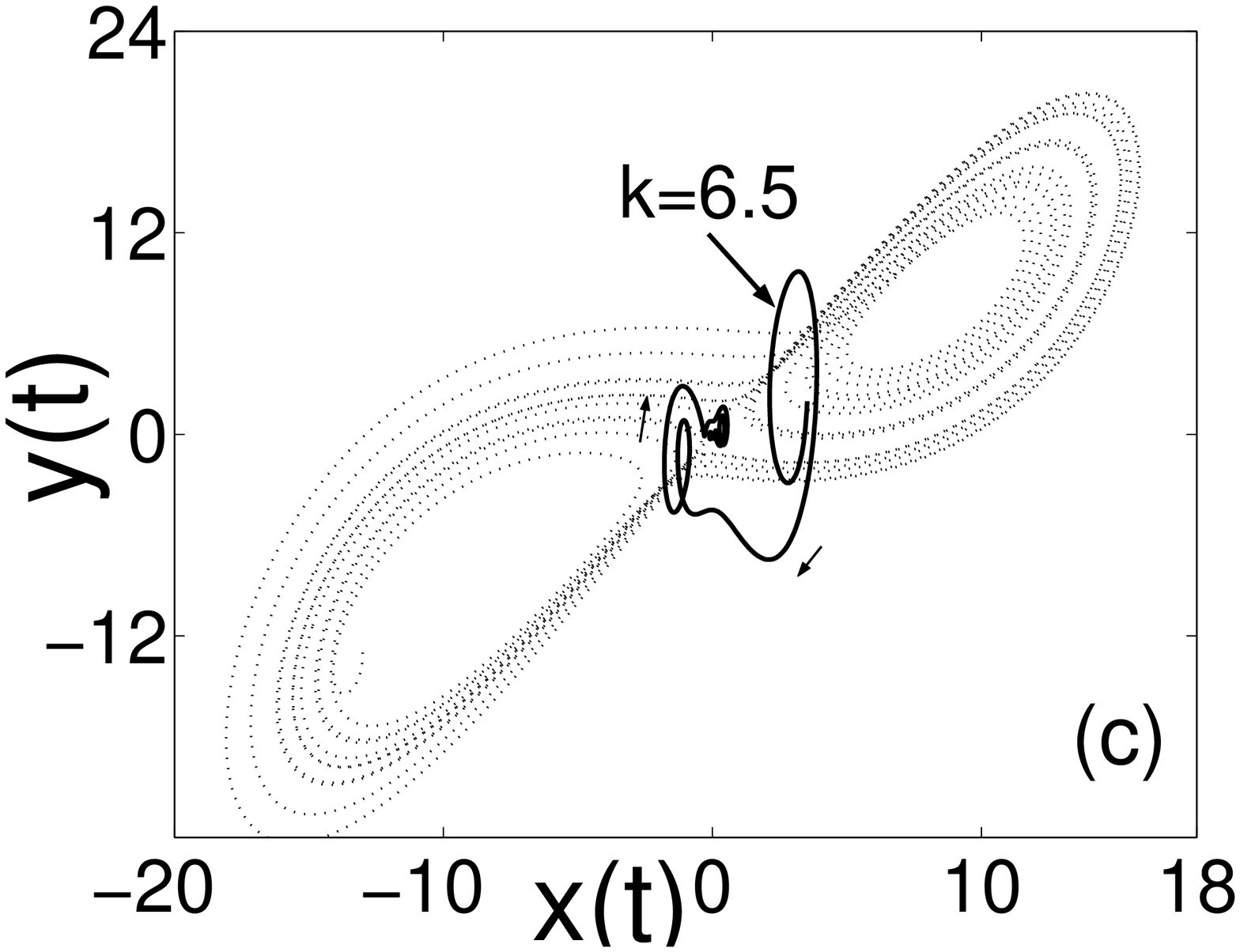}}
\caption{(a) Numerical simulation of chaos suppression shown for
the evolution of the function $x(t)$. The high-frequency
parametric force is turned on at the point $t=20$, and it moves
the system to a new fixed point $<x(t)>=6.8$, for $k=3.5$, or it
stabilizes the origin $x(t)=0$, for $k=6.5$. (b,c) Phase plane
representation of the system dynamics for (b) $k=3.5$ and (c)
$k=6.5$. }
\label{fig_1}
\end{figure}
%%%%%%%%%%%%%%%%%%%%

%%%%%%%%%%%%%%%%%%%%%fig1
\begin{figure}
\centerline{\includegraphics[width=2.7in]{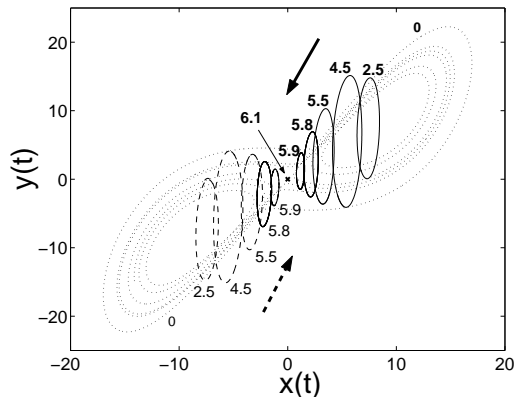}}
\caption{Phase plane representation of the system dynamics after
the suppression of chaos, at different values of the parametric
force amplitude (marked in the figure).} \label{fig_100}
\end{figure}
%%%%%%%%%%%%%%%%%%%%

When the random function $\epsilon(t)$ describes a
bandwidth-limited white noise, $P_{\epsilon}(\omega) = S_0$, the
renormalization factor $K_{\epsilon}$ can be written as
\begin{equation}
K_{\epsilon} = \frac{2\sigma (\omega_2-\omega_1)}{\omega_1
\omega_2} S_0 = \frac{\sigma <\epsilon^2(t)>}{\omega_1\omega_2},
\end{equation}
and the effective control parameter is simplified to be
\begin{equation}
r_{\rm eff} = r \left( 1 - \frac{\sigma r}{\omega_1 \omega_2}
<\epsilon^2(t)>\right),
\end{equation}
so that the chaos suppression effect is clearly proportional to
the noise intensity. Moreover, in the limit $\omega_1, \omega_2
\rightarrow \omega$, we recover the result $K_{\epsilon} = \sigma
k^2/(2\omega^2)$ obtained earlier for the periodic parametric
force [see Eq.~(\ref{eq_r1})].

\section{Numerical simulations and experimental results}

To verify out theory, first we perform direct numerical
simulations of the full model (\ref{eq_1}). Figures~\ref{fig_1}
and \ref{fig_100} show the results for the system temporal
evolution obtained numerically by solving Eqs.~(\ref{eq_1}) and
(\ref{eq_3}) with the parameters: $(\sigma, b, r) = (10, 8/3,
28)$. Since the mean frequency of the unforced oscillations of the
Lorenz system is found to be $\omega_0= 8.24$ [see
Eq.~(\ref{eq_2})], for the above parameters the frequency of the
rapid parametric force is chosen as $\omega=70$. As shown in the
examples presented in Fig.~\ref{fig_1}, the fixed point at the
origin $x=0$ is stabilized for $k = 6.5$ whereas a new stationary
point $X_0 =6.8$ appears for $k = 3.5$. We can see from
Fig.~\ref{fig_100} that the system dynamics in the phase space
after the suppression of chaos is reduced to the coexisting limit
cycles (marked in solid and dotted) whose averaged behavior is
described by Eq.~(\ref{eq_8}), shrinking to the origin as the
parametric force amplitude $k$ is increased.

%%%%%%%%%%%%%%%%%%%%%fig2
\begin{figure}
\centerline{\includegraphics[width=2.8in]{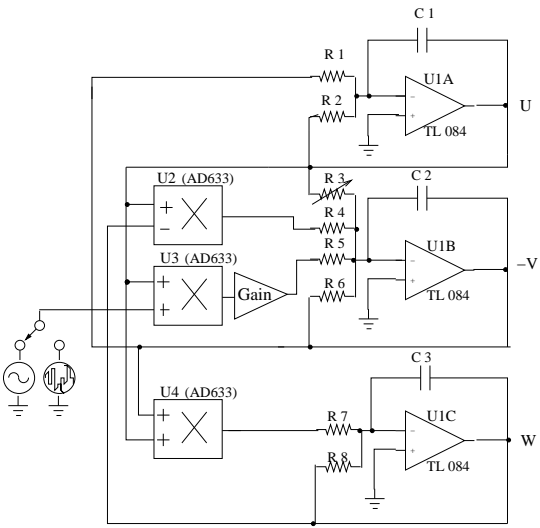}}
\caption{Schematic of the analog electric circuit representing the
Lorenz oscillator in the rescaled variables, including the
parametric driving force and fluctuations. } \label{fig_2}
\end{figure}
%%%%%%%%%%%%%%%%%%%%

Next, we study experimentally the chaos suppression in the Lorenz
system in the framework of an analog electronic circuit
implementation of the Lorenz system (see, e.g.,
Refs.~\cite{cuomo,kurths}). Our circuit uses three Op-Amps (TL084)
as the building blocks for the operations of sum, difference, and
integration, and three analog multipliers (AD633) for the
operations of product.

%%%%%%%%%%%%%%%%%%%%%fig3
\begin{figure}
\centerline{\includegraphics[width=2.7in]{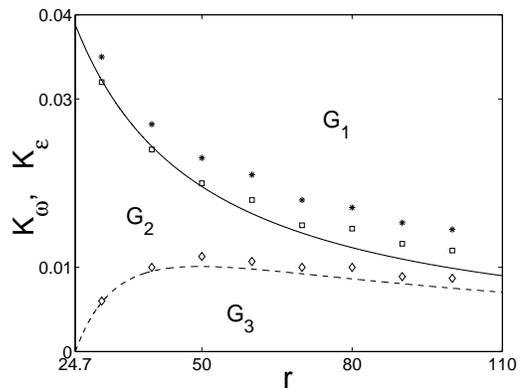}}
\caption{Experimental vs. theoretical results shown as the
stability regions in the parameter plane
($K_{\omega}$/$K_{\epsilon}$, $r$). Solid: theoretical result
[Eq.~(\ref{eq_9}) and Eq.(\ref{eq_18})] for stabilization of the
fixed point at the origin. Dashed: the theoretical results
(\ref{eq_10}) and (\ref{eq_19}) for the transition to a new
oscillating state. Squares and diamonds: experimental results  for
the periodic driving force with the frequency $\omega = 1$KHz.
Asterisks: the critical values $K_{\epsilon}$ for the chaos
suppression by a bandwidth-limited white noise with $\omega_1 =
700$Hz and $\omega_2=1300$Hz. } \label{fig_3}
\end{figure}
%%%%%%%%%%%%%%%%%%%%

As the first step, we rescale the variables $x$, $y$, and $z$ in
order to fit within the dynamical range of the source (-15V, 15V),
and such that the circuit operates in the frequency range of a few
kHz. Applying the transformations $u=x/10$, $v=y/10$, $w=z/10$,
and $\tau = t/A$, we obtain the rescaled Lorenz system, where
$\omega^{\prime} = A\omega$, so that $\omega t =
\omega^{\prime}\tau$ ($\tau$ is measured in seconds). The rescaled
equations describe the temporal evolution of the corresponding
output voltages of Op-Amps in the analog electronic circuit, as
shown schematically in Fig.~\ref{fig_2}.

In the electric circuit shown in Fig.~\ref{fig_2}, the analog
multipliers (AD633) have a noticeable offset at the output that
may alter the dynamical behavior of the system; this effect has
been compensated by using a compensation array. The parameters of
the Lorenz model can be defined through the values of the
resistors, $\sigma = R_6/R_1 = R_6/R_2$, $b=R_6/R_8$, $r=R_6/R_3$,
and the values of the resistors used are: $R_1=R_2 = 100k\Omega$,
$R_3=35.7k\Omega$ (for $r=28$), $R_4=R_7=10k\Omega$,
$R_5=5k\Omega$, and $R_8=374k\Omega$, whereas the values of the
capacitors are: $C_1=C_2=C_3=C = 100nF$. The tolerance of the
resistors and capacitors is less than 1\%. For these values, the
parameters take the following values: $\sigma =10$, $b=2.67$, and
$A=1/R_6C=10(s^{-1})$. The control parameter $r$ of the Lorenz
system and the force amplitude $k$ can be adjusted by varying the
resistor $R_3$ through the relation $r=10^6\Omega/R_3$ and
changing the amplification coefficient of the gain-block element,
respectively, as shown in Fig.~\ref{fig_2}. In the electric
circuit, the frequency of the unforced oscillations for $r=28$ is
found to be $\omega_0^{\prime}=A\omega_0 = 13.1$Hz$= 82.4$ rad/s.
Therefore, we apply a parametric force with the frequency larger
than 65Hz.

%%%%%%%%%%%%%%%%%%%%%fig4

\begin{figure}
\centerline{\includegraphics[width=2.7in]{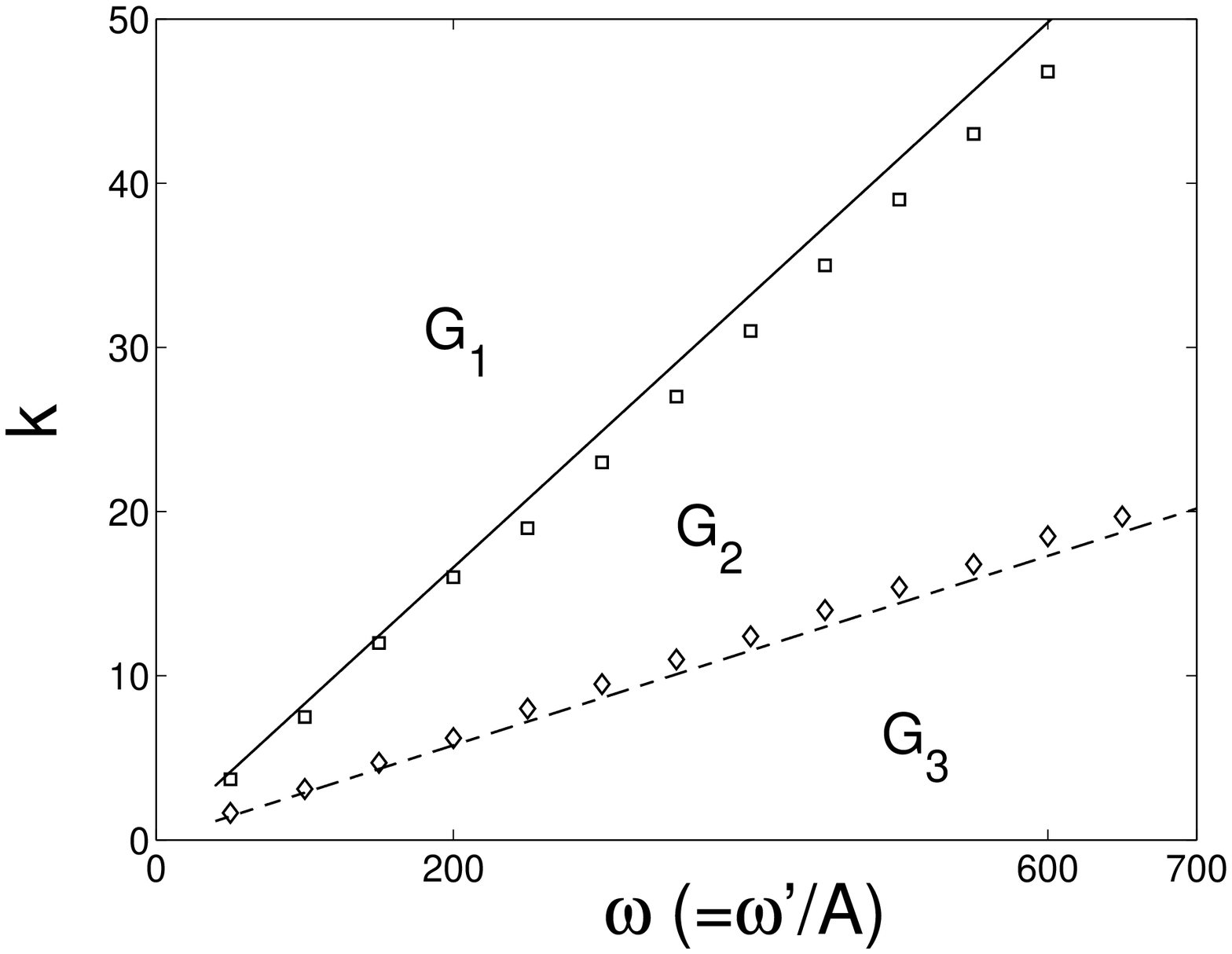}}
\caption{Experimental vs. theoretical results for the critical
amplitude of the parametric force vs. frequency for $r=28$.
Analytical results (\ref{eq_11}) and (\ref{eq_12}) are plotted as
solid and dashed lines, respectively. Squares and diamonds show
the experimental results for the stabilization of the origin and
the suppression of chaos into a fixed point, respectively. }
\label{fig_4}
\end{figure}
%%%%%%%%%%%%%%%%%%%%

Figure~\ref{fig_3} shows the threshold curves described by
Eqs.~(\ref{eq_9}) and (\ref{eq_10}), as well as Eqs.~(\ref{eq_18})
and (\ref{eq_19}) that define three regions with different
nonlinear dynamics, as compared with the numerical and
experimental data. The domains $G_1$ and $G_2$ indicate the
regions for stabilizing the fixed point at the origin and the
transition to a new fixed point, respectively, whereas the domain
$G_3$ is the region of the chaotic dynamics. The critical values
of $K_{\omega}$ were obtained experimentally by changing the
amplitude $k$ of the parametric driving force at the fixed
frequency $\omega = 1$kHz, the corresponding experimental data are
presented in Fig.~\ref{fig_3} by squares and diamonds. In the
experiment with parametric fluctuations (the data are marked by
asterisks), bandwidth-limited white noise with the limits
$\omega_1=700$Hz and $\omega_2=1300$Hz was applied, and its
strength was adjusted by gain-block element (Fig.~\ref{fig_2}).

Figure~\ref{fig_4} shows the dependence of the critical amplitude
of the parametric force for the chaos suppression as a function of
the frequency. Solid and dashed lines represent the theoretical
results described by Eq.~(\ref{eq_11}) and Eq. (\ref{eq_12}),
respectively, whereas the experimental data are plotted as squares
and diamonds. As shown in Fig.~\ref{fig_3} and Fig.~\ref{fig_4},
the experimental results are in a good agreement with the
theoretically calculated threshold dependencies.

\section{Conclusions}

We have studied analytically,  numerically, and experimentally the
suppression of chaos in the Lorenz system driven by a rapidly
oscillating periodic or random parametric force.  We have derived
the theoretical criteria for chaos suppression which indicate
that, for a fixed value of the control parameter $r$, the critical
amplitude of the force required for the suppression of chaos is
proportional to its frequency.  The theoretical criteria for chaos
suppression have been found to agree well with both the results of
numerical simulations and the experimental data obtained for an
analog electronic circuit.

 C.-U.C. and Y.S.K thank the Alexander von Humboldt Stiftung for research
scholarships. The work was also supported by the Deutsche
Forschungsgemeinschaft.

\end{sloppy}
\end{document}